\documentstyle[aps,floats,epsf,twocolumn]{revtex}
\begin{document} \draft

\newcommand{\bk}{{\bf k}}
\newcommand{\bp}{{\bf p}}
\newcommand{\bv}{{\bf v}}
\newcommand{\bq}{{\bf q}}
\newcommand{\br}{{\bf r}}
\newcommand{\bR}{{\bf R}}
\newcommand{\bB}{{\bf B}}
\newcommand{\bA}{{\bf A}}
\newcommand{\bK}{{\bf K}}
\newcommand{\bG}{{\bf G}}
\newcommand{\vd}{{v_\Delta}}
\newcommand{\tvd}{{\tilde v_\Delta}}
\newcommand{\vf}{{v_F}}
\newcommand{\tvf}{{\tilde v_F}}
\newcommand{\TAU}{\mbox{\boldmath $\tau$}}

\title{Universal thermal conductivity in the vortex state of cuprate 
superconductors}

\author{M. Franz$^1$ and O. Vafek$^2$}
\address{$^1$Department of Physics and Astronomy, University of 
British Columbia, Vancouver, BC, Canada V6T 1Z1, \\
$^2$Department of Physics and Astronomy,
Johns Hopkins University, Baltimore, MD 21218
\\
\rm(\today)
\\
~
\\
}
%
\address{~
\parbox{14cm}{\rm We formulate an effective low energy theory for the 
fermionic excitations in $d$-wave superconductors in the
presence of periodic vortex lattices. 
These can be modeled by an effective {\em free}
Dirac Hamiltonian with renormalized velocities and possibly a small
mass term. In the presence of random nonmagnetic impurities
this will result in universal (i.e.\ field and disorder strength independent)
thermal and spin conductivities with values {\em different} from 
those occurring in the Meissner state.
\medskip
}}
\maketitle


At low energies physics of the cuprate superconductors is dominated by
fermionic excitations in the vicinity of the four nodes of the 
$d_{x^2-y^2}$ superconducting order parameter. Formally, these low energy 
excitations  can be described as four species
of relativistic massless Dirac fermions. Perhaps the most spectacular 
manifestation of these Dirac-like excitations is the appearance of universal
conductivities at low temperatures. 
This phenomenon was predicted first by Lee\cite{lee1} for electrical 
conductivity and later extended to spin and thermal conductivities
\cite{graf1,senthil1}.
``Universal'' in this context means that as a result of the linear dispersion 
of Dirac quasiparticles conductivities become independent
of the scattering rate below certain temperature scale (which itself is 
non-universal and set by the scattering rate). Measurement of these universal 
conductivities thus provides information about the intrinsic properties of the 
underlying clean
system. Thermal conductivity has a special significance since its universal
value is unaffected by vertex and Fermi liquid corrections\cite{durst1}.
Experimental measurements of thermal conductivity at sub-Kelvin temperatures
confirmed the existence of the
universal regime in samples of YBa$_2$Cu$_3$O$_{6.9}$ (YBCO)
\cite{taillefer1} and Bi$_2$Sr$_2$CaCu$_2$O$_8$ \cite{chiao1} and provided 
values of the Dirac anisotropy ratio $\alpha_D=v_F/\vd$ in agreement with
other probes, such as the angle-resolved photoemission. To date, no 
observation of universal conductivity in charge or spin channel has been
reported.

In the present paper we argue that, under certain conditions, universal 
conductivities may also appear in the {\em vortex state} of a $d$-wave 
superconductor at low temperatures. In this case ``universal'' means 
independent of {\em both} the quasiparticle scattering rate and the applied 
magnetic field. The physics of this result is based on 
the recent body of work \cite{ft,marinelli1,ye1,vafek1,knapp1,ashvin1,vafek2} 
that resulted 
in detailed understanding of the quasiparticle dynamics 
in the presence of a vortex lattice. Within the simplest linearized
Dirac model the quasiparticles are found to form Bloch bands in the periodic
scalar and vector potentials created by the vortex lattice. One important
conclusion is that the original Dirac node is preserved in the vortex state
\cite{ft,marinelli1} albeit with renormalized quasiparticle
velocities $\tilde v_F$ and $\tvd$. 
Thus, at the lowest energies, one may describe the quasiparticle 
dynamics in the vortex state by an effective {\em free} Dirac Hamiltonian with 
renormalized parameters. Since the $T\to 0$ behavior is entirely determined
by this low energy Hamiltonian, computation of conductivities follows
essentially the same path as in the Meissner state\cite{lee1,graf1,senthil1} 
and yields the aforementioned universal behavior. 

The above sketch captures the essential physics of our result and would
be exact within the linearized model. The reality
is somewhat more complicated. Calculations beyond the linearized 
model\cite{vafek1,ashvin1,vafek2} indicate that internodal scattering and 
nonlinear terms in the full Hamiltonian can modify the band structures
discussed above by displacing or creating additional Dirac nodes or 
producing small gaps at the Dirac point. Such ``massive'' Dirac fermions
have been argued to give rise to quantized thermal Hall conductivity
$\kappa_{xy}$\cite{ashvin1,vafek2}. 
We demonstrate below that, surprisingly, 
even such massive Dirac fermions
give rise to universal {\em longitudinal} conductivities, provided that the 
scattering is in the unitary limit.

Thermal conductivity in the vortex state of a $d$-wave superconductor 
has been addressed previously by number of authors
\cite{kubert1,franz1,vekhter1,vekhter2}. In these works the effect
of vortices has been treated within semiclassical ``Volovik'' approximation
\cite{volovik1} which
tends to capture the essential physics in many situations
\cite{moler1,ong1,chiao2} but  whose domain
of validity remains under debate\cite{marinelli1}. Here, for the first time,
we present a fully quantum treatment of the longitudinal
thermal and spin conductivities in the vortex lattice. Such quantum treatment
is essential when addressing the behavior of very clean systems in the 
$T\to 0$ limit. 

Our starting point is the Bogoliubov-deGennes (BdG)
Hamiltonian linearized near a single node of the $d$-wave order parameter,
\begin{eqnarray}
{\cal H}&=&v_F(\hat p_x +a_x)\sigma_3 + \vd(\hat p_y +a_y)\sigma_1
\nonumber \\
&+& v_Fv_{sx}+ U(\br)\sigma_3.
\label{h0}
\end{eqnarray}
Here $\sigma_i$ are the Pauli matrices, $\hat\bp=-i\nabla$, and we have 
already performed the singular
gauge transformation \cite{ft} to ``unwind'' the phase of the superconducting
order parameter $\Delta=\Delta_0e^{-i\varphi}$. Dirac fermions 
are coupled to the ``Berry'' gauge field ${\bf a}={1\over 2}(\nabla\varphi_A-
\nabla\varphi_B)$, and to the
``Doppler'' gauge field $\bv_s={1\over 2}(\nabla\varphi_A+\nabla\varphi_B
-2e{\bf A}/c)$ which formally enter Hamiltonian (\ref{h0}) 
as vector and scalar potentials respectively. In a static vortex
lattice these two gauge fields are determined through  gradients of 
the phase fields $\varphi_A$ and $\varphi_B$. These represent the singular
parts of the phase coming from two sublattices $A$ and $B$ of the vortex
lattice, and satisfy $\varphi=\varphi_A+\varphi_B$. The last term in 
Eq.\ (\ref{h0}) represents the effects of nonmagnetic impurities. Since 
disorder couples to the electron density, it is invariant under the singular 
gauge transformation. 

\begin{figure}[t]
\epsfxsize=8.5cm
\epsffile{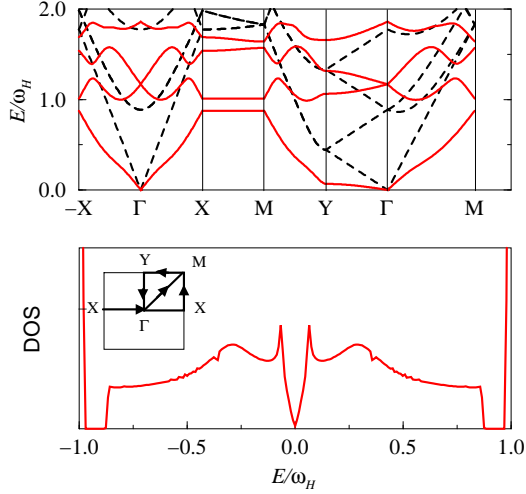}
\caption[]{Low energy band structure for Dirac quasiparticles with 
$\alpha_D=4$ in the presence of square vortex lattice (solid lines) and 
in zero field (dashed lines). Lower panel: associated low energy density 
of states, $\omega_H=\pi\sqrt{\vf\vd(H/2\Phi_0)}$.}
\label{fig1}
\end{figure}

In the absence of disorder and for periodic vortex lattices Hamiltonian
(\ref{h0}) can be diagonalized using standard band structure techniques
\cite{ft,marinelli1,knapp1}. As illustrated in Fig.\ \ref{fig1}, 
for inversion symmetric vortex lattice, spectrum
remains Dirac-like at the lowest energies; the main effect of the vortex 
lattice is to renormalize the Dirac velocities $v_F$ and $\vd$ . 
Once the presence of the nodal points is established \cite{ft,marinelli1,ashvin1}, 
it is straightforward to understand the linear dispersion. 
The wavefunctions can be written in the Bloch form, 
$\Psi_\bk(\br)=
e^{i\bk\cdot\br}\chi_\bk(\br)$, where $\chi_\bk(\br)$ is a two component 
spinor periodic on the unit cell. $\chi_\bk(\br)$ is an eigenstate of the 
Bloch Hamiltonian
$\tilde{\cal H}(\bk)=e^{-i\bk\cdot\br}{\cal H}_0e^{i\bk\cdot\br}$, where 
${\cal H}_0$ is Hamiltonian (\ref{h0}) with the disorder term set to zero. 
At zero energy there is a doublet of degenerate states, 
$|\chi_0^\pm\rangle$, satisfying $\tilde{\cal H}(0)|\chi_0^\pm\rangle=0$.
These zero energy states provide a convenient basis for a perturbative
expansion in the vicinity of the node: we may write
\begin{equation}
\tilde{\cal H}(\bk)=\tilde{\cal H}(0)+(v_F k_x\sigma_3 + \vd k_y\sigma_1),
\label{h1}
\end{equation}
and treat the second term as a small perturbation to $\tilde{\cal H}(0)$ \cite{ashvin1}.
We find that, near the nodal point, a simple first order
degenerate perturbation theory yields very accurate results.
For the energy we obtain 
\begin{equation}
E_\bk^{(1)}=\pm\sqrt{(\tvf k_x)^2+(\tvd k_y)^2},
\label{e1}
\end{equation}
where $\tvf$ and $\tvd$ are renormalized Dirac velocities
to be specified shortly. Higher
order terms produce $O(\bk^2)$ corrections to the energy and are unimportant
at small $\bk$, in agreement with Fig. \ref{fig1}. 

The above considerations imply that very accurate description of the 
low energy excitations of the system in the absence of disorder
can be achieved by projecting its Hamiltonian (\ref{h1}) 
onto the subspace spanned by its two zero energy eigenstates. We now 
include the effects of disorder by projecting
the full Hamiltonian (\ref{h0}) onto this subspace. Formally this is 
accomplished by introducing a projector ${\cal P}=
\sum_{\bk,\nu=\pm}|\bk\nu\rangle\langle\bk\nu|$ with $|\bk,\pm\rangle \equiv
e^{i\bk\cdot\br}|\chi_0^\pm\rangle$. The effective low energy Hamiltonian
${\cal H}^{\rm eff}={\cal PHP}$ reads
\[
{\cal H}^{\rm eff}_{\bk\bk'}=\delta_{\bk\bk'}
\left[\vf k_x({\bf n}_1\cdot{\bf\TAU})
+ \vd k_y ({\bf n}_2\cdot{\bf\TAU})\right]
+{\bf U}_{\bk\bk'}\cdot{\bf\TAU},
\]
where $\tau_i$ are Pauli matrices acting in the $(\mu,\nu)$ space and 
${\bf n}_{1,2}$ are constant vectors defined by
$[{\bf n}_1\cdot{\TAU}]_{\mu\nu}=\langle\bk\mu|\sigma_3|\bk\nu\rangle$ and 
$[{\bf n}_2\cdot{\TAU}]_{\mu\nu}=\langle\bk\mu|\sigma_1
|\bk\nu\rangle$. Similarly ${\bf U}_{\bk\bk'}$ is defined by 
$[{\bf U}_{\bk\bk'}\cdot{\TAU}]_{\mu\nu}=\langle\bk\mu|\sigma_3 U(\br)
|\bk'\nu\rangle$. Making the usual assumption $U(\br)=u_0\sum_\alpha
\delta(\br-\br_\alpha)$ where $\{\br_\alpha\}$ are random impurity positions, 
we obtain
\begin{equation}
[{\bf U}_{\bk\bk'}\cdot{\TAU}]_{\mu\nu}=
u_0\sum_{\alpha,\bG} e^{-i\br_\alpha\cdot(\bk-\bk'-\bG)}
{\cal U}_{\mu\nu}(\bG),
\label{dis1}
\end{equation}
with ${\cal U}_{\mu\nu}(\bG)= 
\langle\chi_0^\mu|\sigma_3e^{-i\br\cdot\bG}|\chi_0^\nu\rangle$. 
Summation
over the reciprocal lattice vectors $\bG$ arises because of the periodicity of
$\chi_0^\mu(\br)$ and will in general 
complicate disorder averaging. However, numerical evaluation of 
${\cal U}_{\mu\nu}(\bG)$ indicates that the sum in Eq.\ (\ref{dis1})
is dominated by the uniform ($\bG=0$) term.
In the following we therefore drop all but this
uniform term, {\em i.e.} ${\cal U}_{\mu\nu}(\bG)\to
\delta_{\bG=0}[{\bf n}_1\cdot\TAU]_{\mu\nu}$. This approximation
allows for simple disorder averaging.

For vortex lattices with inversion symmetry it is easy to show\cite{ashvin1}
that ${\bf n}_1\cdot{\bf n}_2=0$. In such a case, in the absence of disorder,
${\cal H}^{\rm eff}$ has spectrum (\ref{e1}) with $\tvf=\vf|{\bf n}_1|$
and $\tvd=\vd|{\bf n}_2|$. It is then also possible to choose as a basis
such linear combination of the degenerate states $|\chi_0^\pm\rangle$ that
$\vf ({\bf n}_1\cdot{\TAU})=\tvf\tau_3$ and 
$\vd ({\bf n}_2\cdot{\TAU})=\tvd\tau_1$ and write the 
effective low energy Hamiltonian as
\begin{equation}
{\cal H}^{\rm eff}_{\bk\bk'}=\delta_{\bk\bk'}
\left[\tvf k_x\tau_3 + \tvd k_y\tau_1 + m_D\tau_2\right]
+U_{\bk\bk'}\tau_3.
\label{hh3}
\end{equation}
Here $U_{\bk\bk'}={u_0|{\bf n}_1|}\sum_{\alpha} 
e^{-i\br_\alpha\cdot(\bk-\bk')}$ and we added by hand the $m_D$ ``mass'' 
term to model 
the small gap which according to Refs.\ \cite{vafek1,ashvin1} can
open up at the Dirac point as a result of internodal scattering
or nonlinear terms neglected in (\ref{h0}). In the absence of disorder the 
spectrum of Hamiltonian (\ref{hh3}) is 
$E_\bk=\pm\sqrt{(\tvf k_x)^2+(\tvd k_y)^2+m_D^2}$. 

The effective Hamiltonian (\ref{hh3}) is valid at energies $E\ll\omega_H
=\pi\sqrt{\vf\vd(H/2\Phi_0)}$ where $\Phi_0=hc/2e$ is the flux quantum. 
Taking the YBCO values\cite{chiao1} $\vf=2.5\times 10^5$m/s and $\alpha_D=14$ 
we find $\omega_H=25$K$\sqrt{H/1{\rm T}}$: Hamiltonian (\ref{hh3})
should be valid at sub-Kelvin temperatures relevant to the 
low-$T$ heat conduction experiments\cite{taillefer1,chiao1,chiao2}. 

Around the single Dirac node the bare Matsubara Green's function can be written
as
\begin{equation}
\label{greenbare}
\tilde{{\cal G}}_0(\bk,i\omega_n)=
\frac{i\omega_n+\tilde{v}_F k_x\tau_3+\tilde{v}_{\Delta} k_y \tau_1 
+ m_D \tau_2}
{(i\omega_n)^2-(\epsilon_\bk^2+m_D^2)}
\end{equation}
where $\omega_n=(2n+1)\pi T$ and 
$\epsilon_\bk^2=(\tvf k_x)^2+(\tvd k_y)^2$. The impurities alter the 
bare Green's function by introducing a Matsubara self-energy 
$\tilde{\Sigma}(i\omega_n)$.
In the spirit of Ref.\cite{durst1} we assume that all but the scalar component 
of $\tilde{\Sigma}(i\omega_n)$ can be neglected or absorbed into dispersion 
or pairing\cite{pethick1}. Hence the dressed Green's function becomes
\begin{equation}
\label{greendressed}
\tilde{{\cal G}}(\bk,i\omega_n)=\tilde{{\cal G}}_0(\bk,i\omega_n-
\Sigma(i\omega_n)).
\end{equation}
Retarded Green's functions are obtained by analytically continuing 
$\tilde{{\cal G}}_{\rm ret}(\bk,\omega)=
\tilde{{\cal G}}(\bk,i\omega_n \rightarrow \omega+i\delta)$
and the impurity scattering rate is defined as
$\gamma(\omega)=-{\rm Im} \;\Sigma_{\rm ret}(\omega).$

Within the self-consistent t-matrix approximation the self energy
is given by\cite{pethick1,balatsky1}
\begin{equation}
\Sigma(i\omega_n)=\Gamma g_0(i\omega_n)/[c^2-g^2_0(i\omega_n)],
\label{tmatrix}
\end{equation}
with $\Gamma=n_i/\pi\rho_0$, $n_i$ the impurity density, $\rho_0$ the normal
state DOS; $c=\cot \delta_0$ with $\delta_0$ the scattering phase shift, and
$g_0(i\omega_n)=(2\pi\rho_0)^{-1}{\cal N}\sum_\bk{\rm Tr}\tilde{{\cal G}}
(\bk,i\omega_n)$, ${\cal N}$ the number of Dirac nodes. 
Eq.\ (\ref{tmatrix}) self-consistently determines the 
frequency dependent scattering rate $\gamma(\omega)$. At low temperatures
we are interested in $\gamma_0\equiv\gamma(\omega\to 0)$. In the Born limit
$(c\gg 1)$ we find
\begin{equation}
\gamma_0^2\approx -m_D^2 + \Lambda^2e^{-4\pi c^2\tvf\tvd/{\cal N}\Gamma}
\label{born}
\end{equation}
where $\Lambda$ is the upper cutoff of the order of maximum superconducting 
gap. In the massless case this equation always has a real solution for 
$\gamma_0$ implying
finite DOS as $\omega\to 0$ and universal conductivities, albeit below
exponentially small temperatures\cite{lee1}. 
When $m_D>0$ there is no real solution below the critical impurity 
concentration $n_i^c=2\pi^2c^2\tvf\tvd\rho_0/[{\cal N}\ln(\Lambda/m_D)]$: in 
the massive case
weak disorder cannot fill in the gap and produce universal conductivities.

In the unitary limit $(c\to 0)$, we find
\begin{equation}
\gamma_0^2\approx \pi^2 \tvf\tvd\Gamma
\left[{\cal N}\ln{\Lambda^2\over \gamma_0^2+m_D^2}
\right]^{-1}.
\label{unitary}
\end{equation}
This equation has real solution for arbitrarily small impurity concentrations.
In the presence of unitary scattering the system will exhibit finite DOS
at $\omega=0$ and, as we show below, universal conductivities even when 
$m_D>0$. Physically this is a consequence of impurity bound states
forming inside the gap\cite{balatsky1}. Overlap between these states
leads to formation of impurity band capable of carrying the quasiparticle 
current. 

We proceed by computing the spin conductivity $\sigma^s$ 
which is simply related to the heat conductivity by Wiedemann-Franz law
\cite{senthil1,vafek2}, $\kappa/T=(\pi^2k_B^2/3s^2)\sigma^s$.
From now on we shall assume that $\gamma_0>0$, which according to  
the discussion above is guaranteed in the unitary scattering limit.
Neglecting vertex corrections \cite{durst1} 
the static spin conductivity reads
\begin{equation}
\label{sigma}
\sigma^s=\frac{{\cal N}s^2}{2\pi}
\int_{-\infty}^{\infty}d\omega 
\frac{-\partial n_F(\omega)}{\partial \omega} K^s(\omega)
\end{equation}
where $s=1/2$ is the 
coupling constant for  spin current and
\begin{eqnarray}
K^s(\omega)&=&\int \frac{d^2k}{(2\pi)^2}\bigl(
\tvf^2 {\rm Tr}[\tilde{{\cal G}}''_{\rm ret}(\bk,\omega)\tau_3
\tilde{{\cal G}}''_{\rm ret}(\bk,\omega)\tau_3] \nonumber \\
&&\ \ \ \ \ \ \ \ \ \  + 
\tvd^2 {\rm Tr}[\tilde{{\cal G}}''_{\rm ret}(\bk,\omega)\tau_1
\tilde{{\cal G}}''_{\rm ret}(\bk,\omega)\tau_1]\bigr),
\label{kernel}
\end{eqnarray}
with $\tilde{{\cal G}}''_{\rm ret}(\bk,\omega)=
{\rm Im}\tilde{{\cal G}}_{\rm ret}(\bk,\omega)$. In the limit 
$T\rightarrow 0$ we can take 
${-\partial n_F(\omega)/\partial \omega}\rightarrow \delta(\omega)$.
Substituting
\begin{equation}
\tilde{{\cal G}}''_{\rm ret}(\bk,0)=\frac{\gamma_0
-m_D i\tau_2}{\epsilon_\bk^2+m_D^2+\gamma_0^2}
\end{equation}
into Eq.\ (\ref{kernel}) and performing the traces we find
\begin{equation}
K^s(0)={\tvf^2+\tvd^2\over \pi\tvf\tvd} \int_0^\infty \epsilon d\epsilon
{m_D^2+\gamma_0^2\over (\epsilon^2 +m_D^2+\gamma_0^2)^2}.
\label{kernel2}
\end{equation}
The remarkable feature of this result is that the Dirac mass and the scattering
rate enter only in the combination $m_D^2+\gamma_0^2$. This is somewhat 
counterintuitive since in the single particle spectrum the two tend to have 
opposite effects:  mass depletes the low energy DOS while $\gamma_0$ enhances
it. 
 
By power counting the integral in Eq. (\ref{kernel2}) is seen to be
{\em independent} of $m_D^2+\gamma_0^2$, implying universal conductivity just
as in the massless case. We obtain
\begin{equation}
\sigma^s=\frac{{\cal N} s^2}{4\pi^2}\frac{\tvf^2+\tvd^2}{\tvf \tvd}
\end{equation}
and by Wiedemann-Franz law
\begin{equation}
\label{kappa}
\frac{\kappa}{T}=\frac{{\cal N}k^2_B}{12}\frac{\tvf^2+\tvd^2}{\tvf \tvd}\simeq
\frac{{\cal N}k^2_B}{12}\tilde{\alpha}_D,
\end{equation}
where $\tilde{\alpha}_D=\tvf/\tvd\gg 1$ is the renormalized Dirac anisotropy. 
Direct computation of $\kappa$ yields identical result. 

At fields $H_{c1}\ll H \ll H_{c2}$ the only scale in the problem is 
the intervortex separation $l$, implying that to leading order 
$\tilde{\alpha}_D$ 
is field independent\cite{remark2} yet distinct from $H=0$ value. 
Thermal conductivity (\ref{kappa}) is universal: it is 
independent of disorder 
and magnetic field $H$. Numerical calculations within linearized model
\cite{ft,marinelli1,knapp1} show that $\tilde{\alpha}_D>{\alpha_D}$
and thus we expect the quasiparticle contribution to the thermal 
conductivity to be enhanced in the presence of vortices\cite{remark3}.

\begin{figure}[t]
\epsfxsize=7.5cm
\epsffile{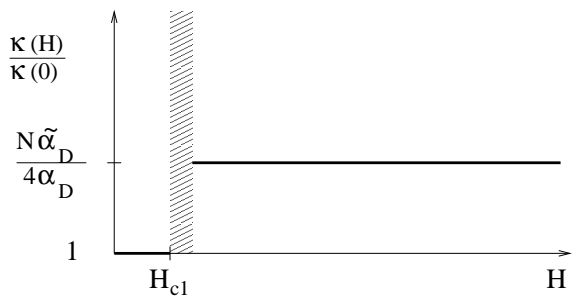}
\vspace{0.5cm}
\caption[]{Schematic plot of the universal quasiparticle thermal conductivity 
$\kappa$ in the vortex state at $T=0$. The contribution from ${\cal N}$ 
species of Dirac 
fermions is proportional to the renormalized Dirac anisotropy 
$\tilde{\alpha}_D > \alpha_D \gg 1$. In the shaded 
region just above $H_{c1}$ the magnetic field is nonuniform
($l\gg\lambda_L$) and our model does not apply. 
}
\label{fig2}
\end{figure}

Fig.\ \ref{fig2} summarizes our result for the field dependence of the 
longitudinal thermal conductivity in the $T\to 0$ limit in the presence
of periodic inversion symmetric vortex lattice. In the Meissner state $(H<H_{c1})$ magnetic field
is excluded from the sample and thermal conductivity assumes its zero field
universal value \cite{graf1} $\kappa_0/T =(k_B^2/3)(\vf/\vd)$. Above $H_{c1}$
formation of the vortex lattice causes renormalization of the Dirac 
velocities and possible gapping or creation of additional nodes. The associated
universal conductivity is $\kappa/T=({\cal N}k_B^2/12)(\tvf/\tvd)$. It is 
worth 
noting that this type of behavior has been recently observed in the ultra-pure
single crystals of YBCO\cite{hill1}.

We conclude by discussing the range of validity of our considerations. 
The crucial
assumption is that of perfectly periodic Bravais vortex lattice with inversion
symmetry. Within the linearized model this guarantees the existence of Dirac
quasiparticles at low energies\cite{marinelli1,ashvin1}. This
assumption should be well satisfied in the ultra-pure single crystals of YBCO. 
Inclusion of internodal scattering and nonlinear effects\cite{vafek1,ashvin1}
neglected in the linearized model, in general can produce small gaps at the 
Dirac points or introduce additional Dirac nodes in the spectrum\cite{vafek2}. 
With the mass term present our effective Hamiltonian (\ref{hh3}) is
sufficiently general to include any such effects. A remarkable result is that,
in the presence of unitary scatterers (or sufficiently high concentration
of Born impurities) even massive Dirac fermions give rise to the universal 
conductivities. We note in passing that this implies universal
conductivities for situations in which the Dirac nodes have been gapped by
other mechanisms, such as in $d_{x^2-y^2}+id_{xy}$ state. This conclusion
holds within the self-consistent t-matrix approximation which corresponds 
to the saddle point analysis in the replicated field theory treatment of
Ref.\ \cite{senthil1}. Ultimately, at the lowest energies, fluctuations around
this saddle point will drive the DOS to zero. It appears, however, that 
this regime has not yet been accessed experimentally, except perhaps in 
samples of underdoped YBCO\cite{behnia1}.

Linearized models indicate that for bare Dirac anisotropies $\alpha_D\gtrsim 
10$ the band structure in the vortex lattice becomes essentially one 
dimensional\cite{ft,marinelli1,knapp1}, i.e. $\tilde\alpha_D\to\infty$. In 
such a case our effective Hamiltonian (\ref{hh3}) would have vanishing
domain of validity. Calculations using the full BdG 
Hamiltonian\cite{vafek1,vafek2} however show that  
the renormalization of the anisotropy is much less severe. 

In less pure samples we expect disorder to destroy periodic vortex lattice.
It has been argued that universal field independent conductivity also emerges
in the random vortex arrays at high fields $H\gg H_{c2}(T/T_c)^2$ 
\cite{franz1,ye1}. This type of behavior was indeed observed at $T\gtrsim 8K$
\cite{ong1}. At sub-Kelvin temperatures, on the other hand, it was found in 
less pure samples that 
$\kappa/T\sim\sqrt{H}$\cite{chiao2}, consistent with semiclassical treatments
\cite{kubert1,vekhter1} which neglect quasiparticle scattering from vortices.
From this perspective our result for clean samples sketched in Fig.\ 
\ref{fig2} represents a nontrivial universal limit with a simple prediction 
that is testable by experiment. Although the size of the jump at $H_{c1}$ 
depends on the details its presence is a robust qualitative prediction 
of the present theory.  

The authors are indebted to R. Hill, C. Lupien and L. Taillefer for discussing
their data on ultra-pure YBCO crystals prior to publication, and are grateful 
to M.-R. Li, A. Melikyan  and Z. Te\v{s}anovi\'{c} for discussions. This work
was supported in part by NSF grant DMR-9415549 (O.V.) and by NSERC (M.F.)


\begin{references}
\bibitem{remark2} We assume that vortex lattice geometry does not change 
with magnetic field.
\bibitem{remark3}  In full nonlinearized model  $\tilde{\alpha}_D$ can 
be smaller than $\alpha_D$.
\end{references}
\end{document}